\documentclass[a4paper]{article}
\usepackage[textwidth=14.9cm]{geometry}
\usepackage{amssymb}
\usepackage[pdftex,dvips]{graphicx}
\usepackage{caption}
\usepackage[]{subcaption}
\usepackage{tabto}
\usepackage[]{array}
\usepackage[]{tabularx}
\usepackage[]{longtable}
\usepackage[]{booktabs}
\usepackage{amsmath}
\usepackage{url}
\usepackage{float}
\usepackage{enumitem}
\usepackage{natbib}
\usepackage{titlesec}
\usepackage{tikz}
\usepackage{pgfplots}
\pgfplotsset{label style={font=\LARGE},tick label style={font=\large}, compat = 1.3}
\newcommand\norm[1]{\left\lVert#1\right\rVert}

\begin{document}

\thispagestyle{empty} 
\renewcommand{\arraystretch}{1.23}
\thispagestyle{empty}

\title{Modeling of Spatial Uncertainties in the Magnetic Reluctivity}
\author{ Radoslav Jankoski$^{1,2}$, Ulrich R\"omer$^{1,2}$ and Sebastian Sch\"ops$^{1,2}$ }
\date{}
\maketitle
{\centering
$^{1}$~Technische Universit\"at Darmstadt, Institut f\"ur Theorie Elektromagnetischer Felder, Darmstadt D-64289, Germany 

$^{2}$~Technische Universit\"at Darmstadt, Graduate School of Computational Engineering, Darmstadt D-64293, Germany 

}

\section*{Abstract}

\textbf{Purpose} - In this paper a computationally efficient approach is suggested for the stochastic modeling of an inhomogeneous reluctivity of magnetic  materials. These materials can be part of  electrical machines, such as a single phase transformer (a benchmark example that is considered in this paper). The approach is based on the Karhunen-Lo{\`e}ve expansion. The stochastic model is further used to study the statistics of the self inductance of the primary coil as a quantity of interest (QoI).

\noindent
\textbf{Design/methodology/approach} - The computation of the Karhunen-Lo{\`e}ve expansion requires solving a generalized eigenvalue problem with dense matrices. The eigenvalues and the eigenfunction are computed by using the Lanczos method that needs only matrix vector multiplications. The complexity of performing matrix vector multiplications with dense matrices is reduced by using hierarchical matrices.

\noindent
\textbf{Findings} - The suggested approach is used to study the impact of the spatial variability in the magnetic reluctivity on the QoI. The statistics of this parameter are influenced by the correlation lengths of the random reluctivity. Both, the mean value and the standard deviation increase as the correlation length of the random reluctivity increases.

\noindent
\textbf{Originality/value} - The Karhunen-Lo{\`e}ve expansion, computed by using hierarchical matrices, is used for uncertainty quantification of low frequency electrical machines as a computationally efficient approach in terms of memory requirement as well as computation time.
\newline
\newline
\noindent
\textbf{Keywords} Karhunen-Lo{\`e}ve expansion, hierarchical matrices, Lanczos method, generalized eigenvalue problem.
\newline
\noindent
\newline
 \textbf{Paper type}: Research paper
\pagebreak
\section{Introduction}
Having an accurate knowledge of the magnetic behaviour law of magnetic materials, expressed via the magnetic reluctivity, plays an important role for optimization of electrical machines. If hysteresis effects are neglected, then there are three main properties related to magnetic materials that need to be taken into account and those are nonlinearity, inhomogeneity and anisotropy. The constitutive relation where all of these properties are included is given as,
\begin{equation}
		\label{reluctivity_introduction}
	\vec{H}(\vec{x})=\overline{\overline{\nu}}(\vec{x},|\vec{B}(\vec{x})|)\vec{B}(\vec{x}),
	\end{equation}
where $\vec{H}$ is the magnetic field strength, $\vec{B}$ is the magnetic flux density and $\overline{\overline{\nu}}$ is the nonlinear, inhomogeneous magnetic reluctivity tensor. However, in practice there is a lack of knowledge of the  magnetic behaviour law, e.g., the manufacturing process introduces variability in this material property. To obtain reliable results from numerical simulations it is necessary to study the impact of those uncertainties. The material properties, that are considered as uncertain, should be modelled as random fields and for the purpose of numerical simulations those random fields must be discretized, i.e., represented via a finite number of random variables. For computationally efficient numerical simulations it is desirable that the number of random variables, through which the random field is represented, is as small as possible. In this sense, we propose to use the Karhunen-Lo{\`e}ve expansion (KLE) to discretize the random reluctivity. When the KLE is truncated after a finite number of terms the truncation error is minimal when compared to any other M-term approximation, e.g., the polynomial chaos expansion, in the mean square sense ~\citep{ghanem1991stochastic}. This property allows to model the random reluctivity with as few random variables as possible. 

In order to use the KLE one needs to know the covariance function of the random field. The covariance function of a spatial random field is usually parametrized by the Mat{\'e}rn family of covariance functions \citep{matérn1986spatial}. In the case when  measurement data is available the parameters of these functions can be obtained by a fitting procedure. In this paper we assume that the covariance function is known and the exponential one is used as a special case of the Mat{\'e}rn family of covariance functions.

So far the KLE has been successfully applied for the stochastic modeling of a nonlinear, homogeneous and isotropic magnetic materials in \citep{doi:10.1137/15M1026535} and the covariance function has been deduced from actual measurements that are already presented in \citep{6208885}. In this paper the KLE is used for the stochastic modeling of spatial uncertainties in linear and isotropic magnetic materials.

A challenge arises within this approach because the KLE requires the solution of a computationally expensive generalized eigenvalue problem with a dense matrix. There are two approaches suggested in the literature in order to remedy this issue. The first approach is described in \citep{Schwab:2006:KAR:1167051.1167057} and the second one in \citep{Khoromskij2009}. Both approaches are based on the Lanczos method that uses only matrix vector multiplications. The first approach consists in accelerating the matrix vector multiplications with the dense matrix by using the fast multipole method (FMM) and the second by using the hierarchical matrix technique. 

The hierarchical matrix technique is adopted in this paper, because it is simpler compared to FMM, and applied to model the magnetic reluctivity of the core of a single phase transformer by using the KLE. The inductance of the primary coil is considered as the quantity of interest(QoI) and the impact of  uncertainty in the spatial distribution of the reluctivity is studied. For many applications it is important to study the mean value and the standard deviation of the QoI.

The main purpose of this paper is to demonstrate the possibility to apply the KLE in combination with the hierarchical matrix technique, as a computationally efficient approach for the stochastic modelling of the magnetic reluctivity, to real life applications such as a single phase transformer. The same approach can be used for any other electrical machine that contains a magnetic material and is  subject to uncertainty.

The work in this paper is organized as follows: In Section 2 the magnetostatic formulation is given and the discretization of the deterministic part by means of the finite element method (FEM). Section 3 introduces details about the KLE. Section 4 gives a brief description of the stochastic collocation method which is used to compute the statistics of the QoI. Section 5 explains the fundamental concepts related to the hierarchical matrix technique. Finally in Section 6 numerical results are given.  

\section {Problem Description}
\subsection{Governing Equation}

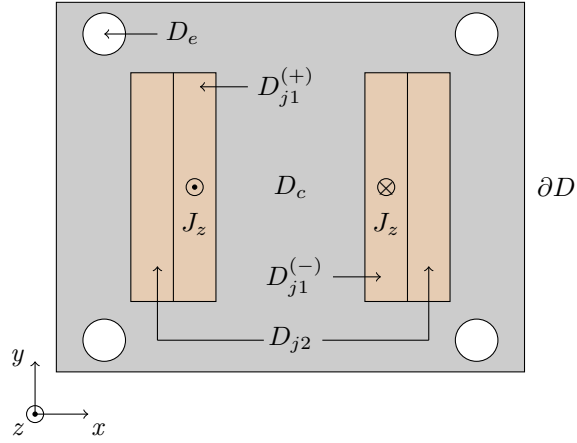
\begin{figure}[t]
\begin{center}
\begin{tikzpicture}[scale=1.4]

\draw [fill= gray!40] (-0.2,0)--(4.2,0)--(4.2,3.5)--(-0.2,3.5) --cycle; 

\draw[fill= brown!40] (0.5,0.6667)--(1.3,0.6667)--(1.3,2.833)--(0.5,2.833)--cycle; 
\draw [fill= brown!40](2.7,0.6667)--(3.5,0.6667)--(3.5,2.833)--(2.7,2.833)--cycle; 

\draw  (3.1,0.6667)--(3.1,2.833);   
\draw  (0.9,0.6667)--(0.9,2.833);    

\draw [fill= white!40](0.25,3.2) circle (0.2);
\draw [fill= white!40](3.75,3.2) circle (0.2);

\draw[fill= white!40](0.25,0.3) circle (0.2);
\draw[fill= white!40] (3.75,0.3) circle (0.2);

\draw [<-] (1.15,2.7)--(1.6,2.7);
\draw [anchor=west] (1.6,2.7) node{$D_{j1}^{(+)}$};

\draw [->] (2.4,0.9)--(2.85,0.9);
\draw [anchor=east] (2.4,0.9) node{$D_{j1}^{(-)}$};

\draw [->] (0.75,0.3)--(0.75,1.0);
\draw (0.75,0.3)--(1.75,0.3);

\draw [->] (3.3,0.3)--(3.3,1.0);
\draw (3.3,0.3)--(2.3,0.3);

\draw  (2,0.3) node{$D_{j2}$};
\draw [<-] (0.25,3.2) -- (0.75,3.2);
\draw [anchor=north west] (0.74,3.4) node{$D_{e}$};

\draw  (2,1.75) node{$D_{c}$};

\draw  (4.5,1.75) node{$\partial{D}$};

\draw (1.1,1.75) circle (0.08);
\draw [fill= black] (1.1,1.75) circle (0.02);
\draw  (1.1,1.4) node{$J_{z}$};

\draw (2.9,1.75) circle (0.08);
\draw  (2.9,1.735) node{$\times$};
\draw  (2.9,1.4) node{$J_{z}$};

\draw (-0.4,-0.4) circle(0.08);
\draw [fill= black] (-0.4,-0.4) circle (0.02);
\draw [->] (-0.4,-0.4) -- (0.1,-0.4);          
\draw [->] (-0.4,-0.4) -- (-0.4,0.1);          

\draw [anchor= north east] (-0.4,-0.4) node{\textit{z}};
\draw [anchor=north west] (0.04,-0.4) node{\textit{x}};
\draw [anchor=north east] (-0.4,0.3) node{\textit{y}};

\end{tikzpicture}

\caption{Single phase transformer that can be found in \citep{FEMM}} 
\label{sin_phase}
\end{center}
\end{figure}

The 2D cross section of the single phase transformer is shown in Fig.\ref{sin_phase}. The computational domain of interest and its boundary are denoted as $D$ and $\partial D$ respectively. The domain where the core of the transformer is located and the air domain (the four air gaps) are denoted as $D_{c}$ and $D_{e}$, respectively. The coil domain $D_{j}$ consists of the primary coil $D_{j1}$ and the secondary coil $D_{j2}$. The regions in the primary coil where the current flows in positive and negative $z$-direction are denoted as $D_{j1}^{(+)}$ and $D_{j1}^{(-)}$, respectively. To compute the self inductance of the primary coil, the secondary coil is left open so that no electric current flows. 

The magnetic reluctivity is defined on each domain separately as follows:
\begin{equation}
\label{reluctivity_def}
 \nu(\vec{x},\theta) =
  \begin{cases}
    \nu_{e} & \text{in $D_{e}$}, \\
    \nu_{j} & \text{in $D_{j}$},  \\
    \nu_{c}(\vec{x},\theta) & \text{in $D_{c}$},
  \end{cases}
  \end{equation}
where $\nu_{e},\nu_{j}$ and $\nu_{c}$ are the magnetic reluctivities of the air, coil and the core domain respectively. The magnetic reluctivity of the core domain $\nu_{c}$ depends on the vector of spatial coordinates $\vec{x}=(x,y) \in \mathbb{R}^{2}$ and and the outcome of a random event $\theta$.

The coil windings usually consist of a considerable number of wires connected in series  and the resolution of every single wire is computationally expensive within FEM. Instead a modeling assumption is introduced: thhe current is assumed to be constant within the cross section of the entire winding and the current density is evaluated as,

\begin{equation}
\label{stranded}
 J_{z} =
  \begin{cases}
    \cfrac{ N_{str}I_{str}}{S_{str}} & \text{in $D_{j1}^{(+)}$}, \\
   \cfrac{  -N_{str}I_{str}}{S_{str}} & \text{in $D_{j1}^{(-)}$}, 
  \end{cases}
\end{equation}
where $N_{str}$ is the number of turns, $I_{str}$ is the current and $S_{str}$ is the surface area of the primary coil winding. This modeling approach is known as stranded conductor model and details can be found in \citep{doi:10.1108/COMPEL-01-2013-0004,250726}.
The behaviour of the system is described with the following simplified 2D stochastic magnetostatic partial differential equation (PDE),
\begin{equation}
	 \label{GOVERNING}
	\begin{aligned}
		-\nabla \cdot (\nu(\vec{x},\theta) \nabla  A_{z}(\vec{x},\theta))=J_z(\vec{x}) \,\,\,\,\,\,\,\, \text{in} \,\,\,\,\,\,\,D,      \\
		 A_{z}(\vec{x},\theta)=0 \,\,\,\,\,\,\,\,\,\,\,\,\,\,\,\,\,\,\,\,\,\,\,\,\,\,\,\,\,\,\,\,\,\,\,\,\,\,\,\, \text{on} \,\,\ \partial D,
		\end{aligned}
	\end{equation}
where $A_{z}$ is the $z$ component of the magnetic vector potential. The normal component of the magnetic flux density is negligible on the boundary $\partial{D}$, hence a Dirichlet boundary condition is imposed in (\ref{GOVERNING}). Equation (\ref{GOVERNING}) is solved by using FEM and details follow in the next subsection.

\subsection{\textit{Finite Element Formulation}}
The PDE (\ref{GOVERNING}) is recast into its weak formulation,
\begin{equation}
	    \label{weak_form}
		  \int_{D}{\nu}(\vec{x},\theta)\nabla  A_{z}(\vec{x},\theta) \cdot \nabla v(\vec{x}) \textnormal{d}{\vec{x}}=\int_{D} J_{z}(\vec{x})v(\vec{x})\textnormal{d}{\vec{x}},
	\end{equation}
where $v$ is a sufficiently smooth test function, subject to the boundary condition in (\ref{GOVERNING}). The solution for the magnetic vector potential $A_{z}$ is approximated with $A_{z}^N$,
\begin{equation}
	    \label{approx}
		A_{z}^N(\vec{x},\theta):=\sum_{i=1}^{N} a_{i}(\theta)  u_{i}(\vec{x}),
	\end{equation}
where $u_{i}$ are globally continuous piecewise linear basis functions on a triangular mesh. Substituting (\ref{approx}) into (\ref{weak_form}) and applying the Galerkin method, results in the system of equations
\begin{equation}
	    \label{fem}
		  \mathbf{K}(\theta) \mathbf{a}(\theta)=\mathbf{F}, 
	\end{equation}
where $\mathbf{K} \in \mathbb{R}^{N \times N}$ is the global stiffness matrix and its elements are calculated as,
\begin{equation}
	    \label{stiffness}
		  {K}_{ij}(\theta):=\int_{D}{\nu}(\vec{x},\theta) \nabla  u_{i}(\vec{x}) \cdot \nabla  u_{j}(\vec{x})\textnormal{d}{\vec{x}}.
	\end{equation}
 The vector $\mathbf{F} \in \mathbb{R}^{N} $ is the loading vector and $\mathbf{a}=(a_{1},..,a_{N})^{T}$ are the coefficients of the magnetic vector potential.

 \subsection{\textit{Inductance Computation}}
The magnetic energy stored in the system is evaluated via the stiffness matrix and the coefficients of the magnetic vector potential as,
 \begin{equation}
	    \label{energy}
		  W(\theta)=\frac{1}{2}\mathbf{a}^{T}(\theta)\mathbf{K}(\theta)\mathbf{a}(\theta),
	\end{equation}
and it is used for the calculation of the inductance
 \begin{equation}
	    \label{Inductance}
		  L(\theta)=\frac{2 W(\theta)}{I_{str}^{2}}.
	\end{equation}
In the next section we replace the dependency on $\theta$ with a finite dimensional vector of random variables by means of the KLE.

\section{Karhunen-Lo{\`e}ve Expansion}

The magnetic reluctivity is expressed via a finite number of random variables by the truncated KLE as follows:
	\begin{equation}
		\label{cs}
		{\nu}_{c}(\vec{x},\theta) \approx  \overline{\nu}(\vec{x})+\sum_{i=1}^{M} \sqrt{\lambda_{i}}f_{i}(\vec{x})\xi_{i}(\theta),
	\end{equation}
where $\overline{\nu}$ is the mean value of the random field, $\vec{\xi}$ is a vector of mutually uncorrelated orthonormal random variables, $f_{i}$ are orthonormal eigenfunctions and $\lambda_{i}$ are eigenvalues. A comprehensive theoretical treatment of the KLE given with equation (\ref{cs}) can be found in \citep[chapter 4, p.~47]{Xiu:2010:NMS:1893088} and \citep[chapter 2, p.~18]{maitre2010spectral}. The truncation parameter $M$ is chosen such that $\Psi_{M}>0.95$ (also known as relative information criterion), where
\begin{equation}
	    \label{truncation}
		  \Psi_{M}=\frac{\sum_{i=1}^M\lambda_{i}}{\sum_{i=1}^{\infty}\lambda_{i}}.
	\end{equation}
The eigenfunctions and the eigenvalues that appear in the KLE are obtained by solving the Fredholm integral equation:
	\begin{equation}
	    \label{fredholm}
		  \int_{D_{c}}\textnormal {Cov}(\vec{x},{\vec{y}}) f_i(\vec{x})\textnormal{d}{\vec{x}}={\lambda_i}f_i(\vec{y}),
	\end{equation}
where $\textnormal {Cov}$ is the covariance function. The air and the coil domain are excluded from equation (\ref{fredholm}) because the covariance function is nonzero only where the core is located (the reluctivity of the air and the coil is considered to be deterministic). Equation (\ref{fredholm}) has an analytical solution for some particular covariance functions such as the exponential one \citep[Chapter 4, p.~48]{Xiu:2010:NMS:1893088} defined on a rectangular domain. However in the general case it has to be solved numerically. For that purpose the eigenfunction is approximated as follows:
  \begin{equation}
	    \label{aprxx}
		 f_i(\vec{x})\approx \sum_{j=1}^{P}f_{ij} \phi_{j}(\vec{x}),
	\end{equation}
where $\phi_{j}$ are shape functions to be specified. If (\ref{aprxx}) is substituted in (\ref{fredholm}) and both sides are multiplied with the test function $\psi_{j}$ and integrated over the core domain, the following generalized eigenvalue problem is obtained,
	\begin{equation}
	\label{gem}
\mathbf{A}\mathbf{f}={\lambda}{\mathbf{B}}{\mathbf{f}},
\end{equation}
with matrices $\mathbf{A},\mathbf{B} \in \mathbb{R}^{P \times P}$ and eigenvector $\mathbf{f}\in \mathbb{R}^{P}$. The elements of the matrix $\mathbf{A}$ are calculated as follows:
\begin{equation}
	    \label{AIJ}
		 A_{ij}:= \int_{D_{c}}\int_{D_{c}}\textnormal {Cov}(\vec{x},{\vec{y}}) \phi_{i}(\vec{x})\psi_{j}({\vec{y}})\textnormal{d}{\vec{x}}\textnormal{d}{\vec{y}},
	\end{equation}
and the elements of the matrix $\mathbf{B}$ are given by
\begin{equation}
	    \label{BIJ}
		 B_{ij}:=\int_{D_{c}} \phi_{i}(\vec{y})\psi_{j}({\vec{y}})\textnormal{d}{\vec{y}}.
	\end{equation}
The same mesh, which is used for the FEM part, is used to solve the Fredholm equation. Since the FEM solution consists of piecewise linear basis functions, only one Gauss point is required for the numerical integration of stiffness matrix entries. Hence, no higher order approximation for $f$ is required and constant basis and shape functions are chosen. As a consequence $P=N_{t}$, where $N_{t}$ is the number of triangles in the mesh. Matrix $\mathbf{A}$ is symmetric and dense and $\mathbf{B}$ is a diagonal matrix.

The random variable that appears in the KLE $\xi_{i}$ is evaluated as follows:

\begin{equation}	
\label{rv}	
		\xi_{i}(\theta)=\frac{1}{\sqrt{\lambda_{i}}}\int_{D_{c}}({\nu}(\vec{x},\theta)-\overline{\nu}(\vec{x})){f_{i}}(\vec{x})\textnormal{d}{\vec{x}}.
	\end{equation}
Equation (\ref{rv})	states that in order to determine the probability density function (pdf) random realizations from the magnetic reluctivity  are needed. In practice it means that the pdf is determined from actual measurements. In this paper a uniform pdf of $\xi_{i}$ is assumed.
By using equation (\ref{cs}) the stiffness matrix can be written as follows:
\begin{equation}	
\label{sts}	
		\mathbf{K}(\vec{\xi})=\overline{\mathbf{K}}+\sum_{i=1}^{M} \mathbf{K}_{i} \xi_{i}(\theta).
	\end{equation}
The KLE separates the deterministic part from the random part of the random field. This property is computationally useful because the stiffness matrices $\{\mathbf{K}_{i}\}^{M}_{i=1}$ are not recomputed for every new sample of the random variable $\xi_i$, as it can be seen from equation (\ref{sts}). The mean value of the inductance is evaluated as follows:
\begin{equation}
\label{m_value}
 L_{\mu}=\int_{\Xi}L(\vec{\xi})\rho(\vec{\xi})\textnormal{d}{\vec{\xi}},
\end{equation}
and its standard deviation,
\begin{equation}
\label{vsr_value}
 L_{std}=\sqrt{\int_{\Xi}(L(\vec{\xi})-L_{\mu})^{2}\rho(\vec{\xi})\textnormal{d}{\vec{\xi}}},
\end{equation}
where $\rho$ is the joint probability function of the random variables $\{\xi_{i}\}^{M}_{i=1}$ and $\Xi=[a_{1},b_{1}] \times [a_{2},b_{2}]\times .. \times [a_{M},b_{M}]$ where the interval $[a_{i},b_{i}]$ is the support of the pdf of the random variable $\xi_{i}$.

\section{Stochastic Collocation Method}

To compute the statistics of the inductance, the stochastic collocation method is used \citep{doi:10.1137/040615201,doi:10.1137/050645142}. The FEM solution is computed at the full tensor grid of multidimensional collocation points $\{\vec{\xi_{k}}\}^{N_{c}}_{k=1}$, obtained as tensor product of one dimensional Gauss-Legendre collocation points, where $N_{c}$ is the total number of points. Each collocation point $\vec{\xi_k}=(\xi_{k,1},...,\xi_{k,M})$ consists of $M$ components. This results is $N_{c}$ decoupled system of deterministic equations and the discrete solution is then interpolated as follows:
\begin{equation}
\label{interpolation}
L(\vec{\xi})\approx \sum_{k=1}^{N_{c}}L(\vec{\xi_{k}})h_{k}(\vec{\xi}),
\end{equation}
where $h_{k}$ is a multivariate Lagrange polynomial. The degree of the one dimensional Lagrange polynomials is denoted as $p$. The mean value and the standard deviation are approximated by a numerical quadrature
\begin{equation}
\label{m_value_coll}
 L_{\mu}\approx \sum_{k=1}^{N_{c}}L(\vec{\xi_{k}})w_{k},
\end{equation} 
and,
\begin{equation}
\label{std_value_coll}
 L_{std} \approx \sqrt{\sum_{k=1}^{N_{c}}(L_{\mu}-L(\vec{\xi_{k}}))^{2}w_{k}},
\end{equation} 
where $w_{k}$ is the weight coefficient, respectively. 

\section{Hierarchical Matrices}
The basic idea behind the hierarchical matrix technique is to find certain subblocks in the dense matrix $\mathbf{A}$, that have small entries (because they are far away from each other and weakly coupled) and perform a low-rank approximation. The low rank approximation of a dense matrix decreases the computational costs of the basic arithmetical operations such as addition, matrix-matrix multiplication and matrix-vector multiplication and also reduces the memory storage requirements. The key elements of building a hierarchical matrix representation of a dense matrix are a cluster tree and a block cluster tree. The indices of the potential candidates (subblocks) for a low rank approximation are stored in a block cluster tree whose elements are obtained as a Cartesian product between the elements of a cluster tree. Fundamental theory behind the hierarchical matrices can be found in \citep{HA99,HAKHSA00,BOHA02} and \citep{GRHA02, HA09, börm2010efficient}. For the sake of clarity we give a brief description about the cluster and block cluster tree and the low rank approximation technique.
\subsection{\textit{Cluster Tree}}
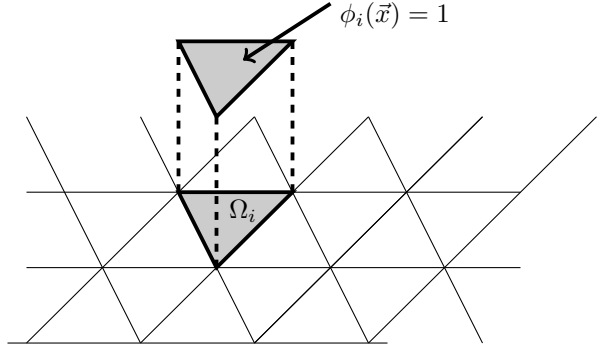
\begin{figure}[t]

\begin{center}
\begin{tikzpicture}[scale=0.5]

\draw (-0.5,0)--(9.5,0);  
\draw(0,0)--(6,6);

\draw(3,0)--(9,6);
\draw(6,0)--(12,6);

\draw(6,0)--(12,6);
\draw(9,0)--(15,6);

\draw (0,2)--(13,2);

\draw(3,0)--(0,6);
\draw(6,0)--(3,6);
\draw(9,0)--(6,6);
\draw(12,0)--(9,6);

\draw (0,4)--(13,4);   
\draw [fill= gray!40,line width=0.5mm] (4,4) -- (7,4) -- (5,2) --cycle; 

\draw[dashed,line width=0.5mm] (5,2)--(5,6);
\draw[dashed,line width=0.5mm] (7,4)--(7,8);
\draw[dashed,line width=0.5mm] (4,4)--(4,8);

\draw [fill= gray!40,line width=0.5mm] (5,6) -- (7,8) -- (4,8) --cycle; 

\draw (5.7,3.5) node{ $\Omega_{i}$};

\draw [<-,line width=0.5mm]((5.7,7.5) -- (8,9); 
\draw [anchor=north west] (8,9.3) node{ $\phi_{i}(\vec{x})=1$};

\end{tikzpicture}

\caption{Support of a piecewise constant basis function} 
\label{support}
\end{center}
\end{figure}

Let $I$ be an index set $I=\{0,1,2.. .N_{t}-1\}$. Each element $i \in I$ references a domain $\Omega_{i}$ described by the following expression
\begin{equation}
	\label{supp}
\Omega_{i}:=\textnormal{supp}(\phi_{i}).
\end{equation}
A piecewise constant basis functions is defined as follows:
\begin{equation}
\label{supp_phi}
  \phi_i(\vec{x}) :=
  \begin{cases}
    1 & \text{if $\vec{x} \in \tau_{i}$} \\
    0 & \text{otherwise}
  \end{cases}
 \end{equation} 
From equation (\ref{supp_phi}) it is concluded that the domain $\Omega_{i}$ corresponds to the domain of $\tau_{i}$, the $i-\textnormal{th}$ triangle in the triangular mesh, as it is illustrated in Fig.(\ref{support}).

The tree $T_{I}$ is called a \textbf{cluster tree} over the index set $I$ if the following conditions hold:
\begin{itemize}
\item The index set $I$ is the root of the cluster tree.
\item If $t \in T_{I}$ is not a leaf, then it is a disjoint union of its sons $S(t)$.
\item If $t \in T_{I}$ is a leaf, then $\#t \leq n_{min}$ for fixed number $n_{min}$.  
\end{itemize}
The operation $\#$ refers to the cardinality of a set. There are many different clustering algorithms to build a cluster tree, for instance algebraic, geometrical, cardinality-balanced and box-tree clustering algorithm \citep{JD}. Regardless of the type of the algorithm the basic idea is to split the index set $I$ into two disjoint subsets which become sons of the root cluster. This procedure is repeated recursively for the son clusters. In this paper the box-tree clustering algorithm is used  because of its advantages over the other algorithms that are explained in \citep[Chapter 2, p.~30]{JD}.

The clusters $t \in T_{I}$ define cluster domains $\Omega_{t}$,
\begin{equation}
	\label{cluster_domain}
\Omega_{t}:=\bigcup\limits_{i \in t}\textnormal{supp}(\phi_{i}),
\end{equation}
i.e., $\Omega_{t}$ is the minimal subset of $\mathbb{R}^{2}$ that contains the supports of the basis functions $\{\phi_{i}\}_{i \in t}$. The potential candidates for low rank approximation are those subblocks in the matrix $\mathbf{A}$ whose indices are stored in the clusters $t,s \in T_{I}$ and their domains $\Omega_t$ and $\Omega_s$ defined with equation (\ref{cluster_domain}) satisfy the so called admissibility condition,
\begin{equation}
	\label{adms}
      \textnormal{min}(\textnormal{diam}(\Omega_t),\textnormal{diam}(\Omega_s))\leq \eta \textnormal{dist}(\Omega_t,\Omega_s),
\end{equation}
where $\textnormal{diam($\cdot$)}$ is the Euclidean diameter of a set and $\textnormal{dist($\cdot,\cdot$)}$ is the Euclidean distance of two sets. The parameter $\eta$ allows to adjust the number of admissible blocks.

\begin{figure}[t]

\begin{center}
\begin{tikzpicture}[scale=0.55]

\draw [fill= gray!10] (0,0) rectangle (4.5,4.5); 

\draw [fill= gray!40] (0.5,0) -- (2,2) -- (0,2) --cycle;
\draw [fill=gray!40] (0,2) -- (1.5,4.5) --(2,2) --cycle;
\draw [fill= gray!40] (0.5,0) -- (4.5,1.5) -- (2,2) --cycle;
\draw [fill= gray!40] (1.5,4.5) -- (4.5,1.5) -- (2,2)--cycle;
\draw [fill= gray!40] (1.5,4.5) -- (4.5,1.5) -- (4,4)--cycle;

\draw [line width = 0.9mm,red,<->] (0,0) -- (4.5,4.5); 
\draw [dashed,line width = 0.7mm,blue,<-] (4.1,4.1) -- (5.5,3.5); 
\draw [anchor=north west](5.1,3.6) node{ $\textnormal{diam}(Q_t)$}; 

\draw [fill= gray!10](10,6) rectangle (14.5,10.5);  

\draw [fill= gray!40] (11.2,6) -- (10,8.5) -- (12.4,8.2) --cycle;

\draw [fill= gray!40] (10,8.5) -- (11.4,10.5) -- (12.4,8.2) --cycle;

\draw [fill= gray!40] (10,8.5) -- (11.4,10.5) -- (12.4,8.2) --cycle;

\draw [fill= gray!40] (11.4,10.5) -- (12.4,8.2) -- (14.5,9.2) --cycle;

\draw [fill= gray!40] (11.2,6) -- (14.5,9.2) -- (12.4,8.2) --cycle;

\draw [line width = 0.9mm,red,<->] (10,6) -- (14.5,10.5); 

\draw [line width=0.7mm ,green, <->] (4.5,4.5)--(10,6); 

\draw (13.5,7.5) node{ $Q_s$}; 

\draw (3.2,0.5) node{ $Q_t$};

\draw (7.5,4.5) node{ $\textnormal{dist}(Q_t,Q_s)$}; 
\draw [dashed,line width = 0.7mm,blue,<-] (7.0,5.1) -- (7.6,4.8); 

\draw (11.7,5.1) node{ $\textnormal{diam}(Q_s)$}; 
\draw [dashed,line width = 0.7mm,blue,<-] (10.35,6.3) -- (11.5,5.3);

\end{tikzpicture}
\end{center}

\caption{Example of two separated cluster domains}
     \label{admiss}

\end{figure}
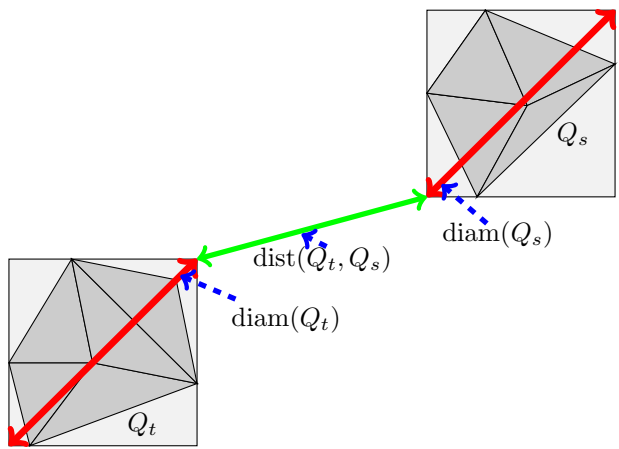

In practice, computing the diameters of the possibly complicated domains $\Omega_t$ and $\Omega_s$, as well the distance between them, can be a time consuming procedure. Thus they are replaced by axis-parallel rectangles $Q_{t}$ and $Q_{s}$ such that $\Omega_{t} \subseteq Q_{t}$ and $\Omega_{s} \subseteq Q_{s}$ holds as it is depicted in Fig.(\ref{admiss})

The admissibility condition is examined as follows:
\begin{equation}
	\label{adms_rec}
      \textnormal{min}(\textnormal{diam}(Q_t),\textnormal{diam}(Q_s))\leq \eta \textnormal{dist}(Q_t,Q_s),
\end{equation}
as an alternative to equation (\ref{adms}). The admissibility condition given with equation (\ref{adms}) as well as (\ref{adms_rec}) is tailored to kernel functions that have singularities at $\vec{x}=\vec{y}$. In the case studied here the kernel function is the covariance function and it is not singular at $\vec{x}=\vec{y}$ and yet the same admissibility condition is used. The question of finding an admissibility condition which gives an optimal hierarchical matrix representation for non singular kernels is still open. A detailed discussion on this topic can be found in \citep{Khoromskij2009}.

\subsection{\textit{Block Cluster Tree}}

The tree $T_{I \times I}$ is called a \textbf{block cluster tree} over the Cartesian product $I \times I$ if the following holds:
\begin{itemize}
\item 
The root of the block cluster tree is $I \times I$. 
\item
Each element $b \in T_{I \times I} $ has the form $b=t \times s$ where $t,s \in T_{I}$.
\end{itemize}
The block cluster tree is built by taking a Cartesian product between the elements $t,s \in T_{I}$ that belong to the same level in the cluster tree $T_{I}$ and the procedure is repeated recursively until the admissibility condition is satisfied or $t$ and $s$ are leaves. The block cluster tree is a quadtree with leaves on different levels that could be admissible or inadmissible. A hierarchical matrix represents also a cluster tree with an equivalent structure as the block cluster tree. The inadmissible leaves of the hierarchical matrices contain matrices in the standard full format and the admissible leaves contain low rank approximated matrices. 
\subsection{\textit{Low Rank Approximation}}
The matrix subblock $\mathbf{A}_{t \times s}$ is low rank approximated by the matrix subblock $\mathbf{\tilde{A}}_{t \times s}$, if the clusters $t$ and $s$ satisfy the admissibility condition, in the following way:
\begin{equation}
	\label{lr}
\mathbf{\tilde{A}}_{t \times s}=\mathbf{R}\mathbf{K}^{T},
\end{equation}
where $\mathbf{R} \in \mathbb{R}^{p \times k}$, $\mathbf{K} \in \mathbb{R}^{q \times k}$ and  $p=\#t$, $q=\#s$. The rank is denoted as $k$. In this paper the matrices $\mathbf{R}$ and $\mathbf{K}$ are computed by using the adaptive cross approximation (ACA) technique \citep{Bebendorf2000}. ACA computes a low rank approximation  with desired accuracy $\varepsilon$ such that
\begin{equation}
	\label{eps}
\norm{\mathbf{A}_{t \times s}-\mathbf{\tilde{A}}_{t \times s}}_{F}\leq\varepsilon\left\lVert\mathbf{A}_{t \times s}\right\rVert_{F},
\end{equation}
where $\norm{\cdot}_{F}$ is the Frobenius norm. 
  
\pagebreak

\section{Results}
\subsection{Computation of Eigenvalues and Eigenfunctions}
The domain of interest $D$ is triangulated by the open source mesh generator triangle \citep{TRIANGLE} within the FEMM 4.2 software \citep{FEMM_ONLINE_ONE}. The stiffness matrix given with equation (\ref{stiffness}) is computed and assembled by using an inhouse developed MATLAB code Niobe. For computing and storing the matrix $\mathbf{A}$ into a hierarchical format we use the H2Lib library \citep{h2lib}. For the eigenvalue and eigenfunction computations we use our own implemented Lanczos solver.

\begin{figure}[b]

\captionsetup[subfigure]{justification=centering}
\centering
\begin{tikzpicture}[scale=0.75,trim axis left, trim axis right]
     
     \begin{semilogyaxis}[
        xlabel=\Large  $i$,
        ylabel=\Large $\lambda_{i}$, xlabel shift=0 pt,ylabel shift=0 pt,legend cell align=left
                         ]
  
      \addplot table{eigendata_d002.txt};
      \addplot table{eigendata_d010.txt};
      \addplot table{eigendata_d100.txt};
            
      \legend{\large $d=2m$\\\large  $d=10m$\\\large  $d=100m$\\}
    \end{semilogyaxis}
 
      \end{tikzpicture}

      \caption{Eigenvalues}
         \label{eigenvalues}
  \end{figure}
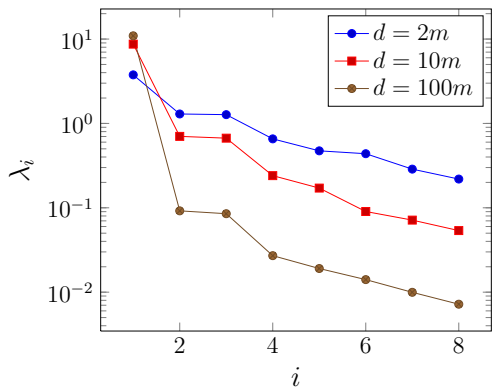

For illustration  purposes the Fredholm integral equation is solved on the domain $D_{c}$ for a random field with a given analytical covariance function instead of deducing it from measurements,
\begin{equation}
\label{exponential}
\textnormal {Cov}(\vec{x},{\vec{y}})=\sigma^{2}\exp{\Bigg( -\frac{||\vec{x}-\vec{y}||_{l_{1}}}{d}\Bigg)},
\end{equation}
where $\sigma$ is the standard deviation of the random field and the correlation length is denoted as $d$. The number of triangles in the mesh is $N_{t}=24727$. The parameters related to the hierarchical matrix representation are $n_{min}=256$, $\eta=1.0$ and $\varepsilon=0.01$. For the numerical integration of equation (\ref{AIJ}) only one quadrature per element yields a sufficient accuracy. The eigenvalues are shown  for three different correlation lengths in Fig.(\ref{eigenvalues}) for $\sigma=1$. As one would expect, the decay of the eigenvalues depends on the correlation length. For strongly correlated random fields the decay is quite fast (orders of magnitudes). 
		
\begin{figure}[ht]
   \centering
    \begin{subfigure}{7.3cm}
     \centering 
      \includegraphics[width=7.2cm]{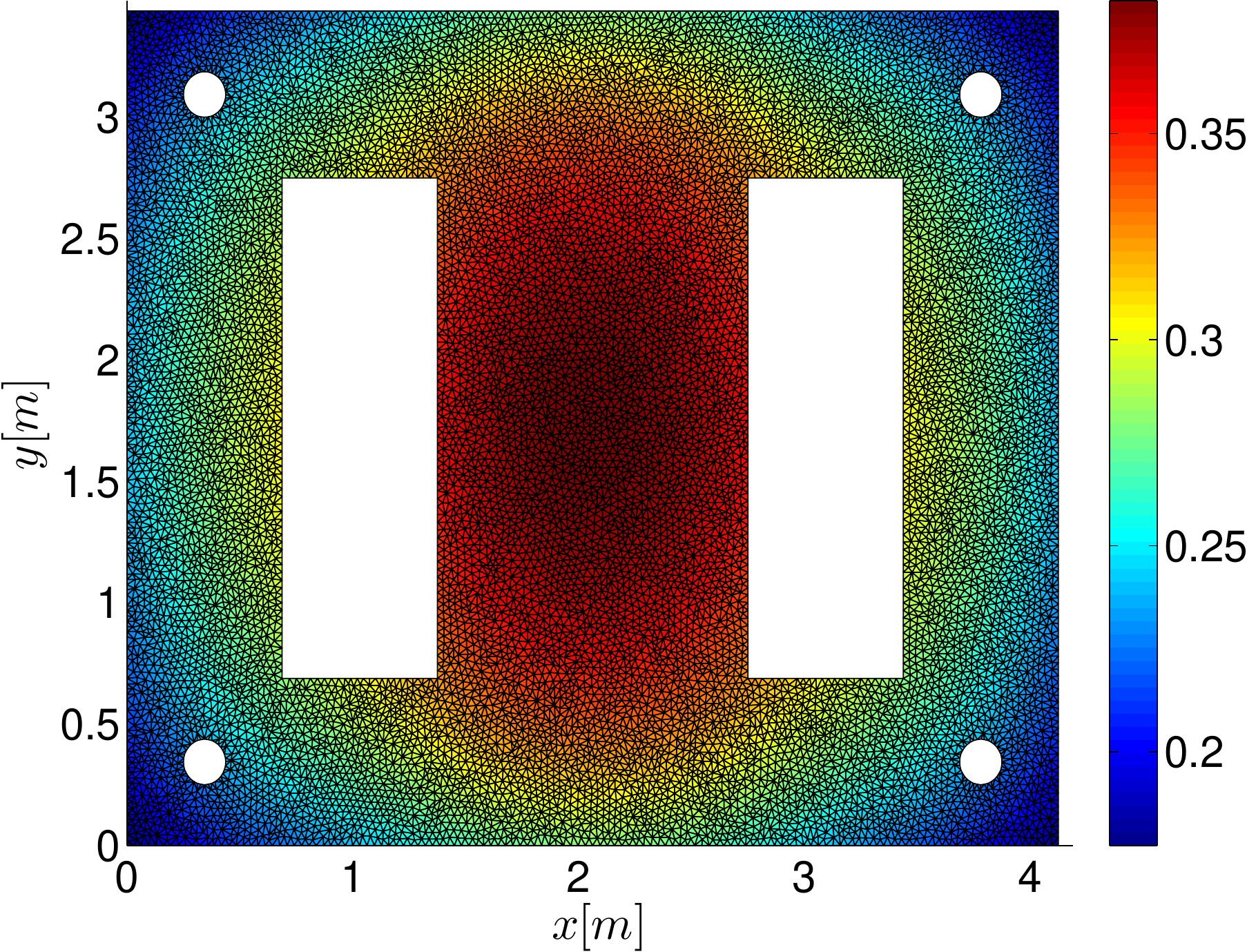}
       \caption{1st eigenfunction}
    \end{subfigure}
\begin{subfigure}{7.3cm}
  \centering
      \includegraphics[width=7.2cm]{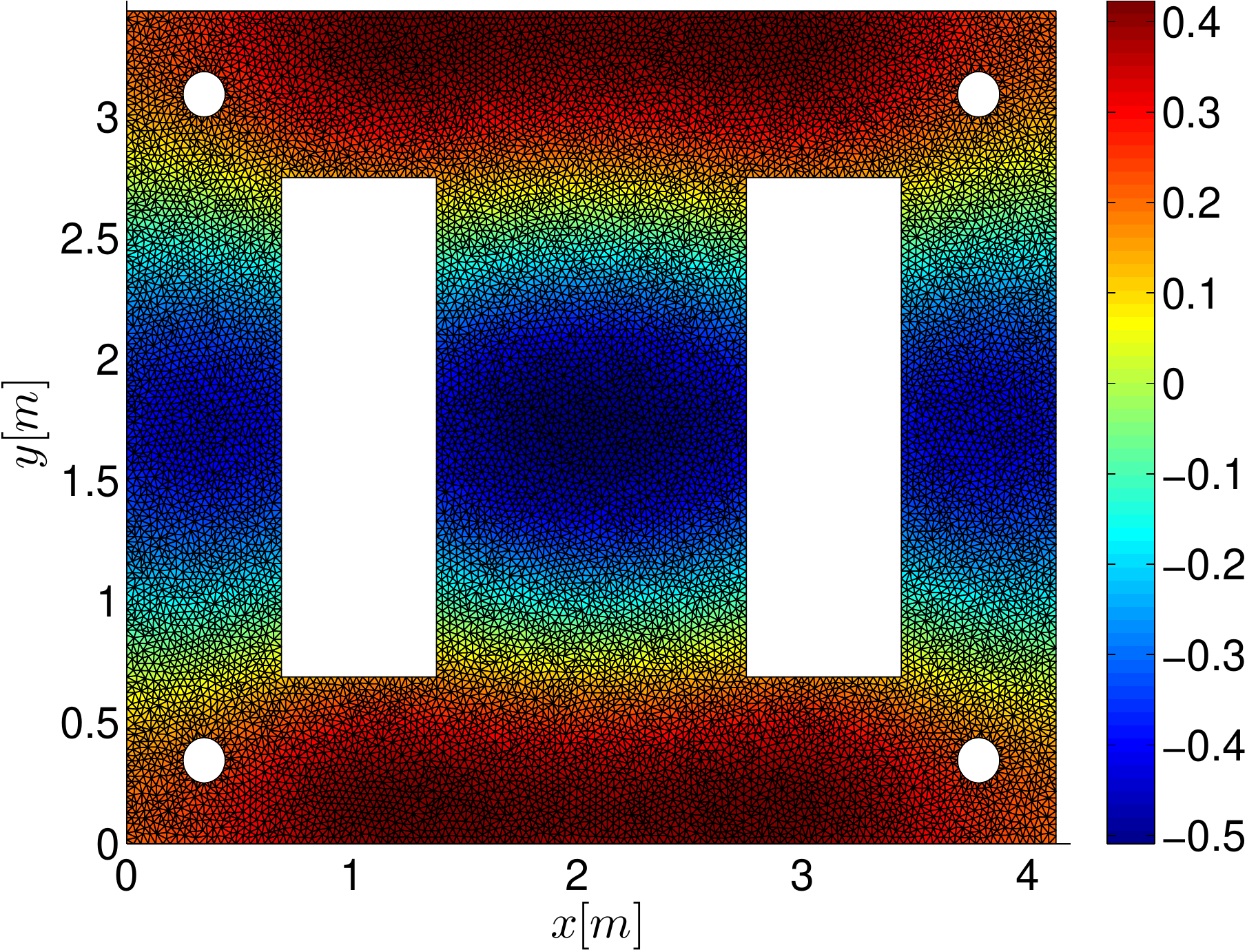}
       \caption{5th eigenfunction}
    \end{subfigure}
   \centering
    \begin{subfigure}[b]{7.3cm}
      \includegraphics[width=7.2cm]{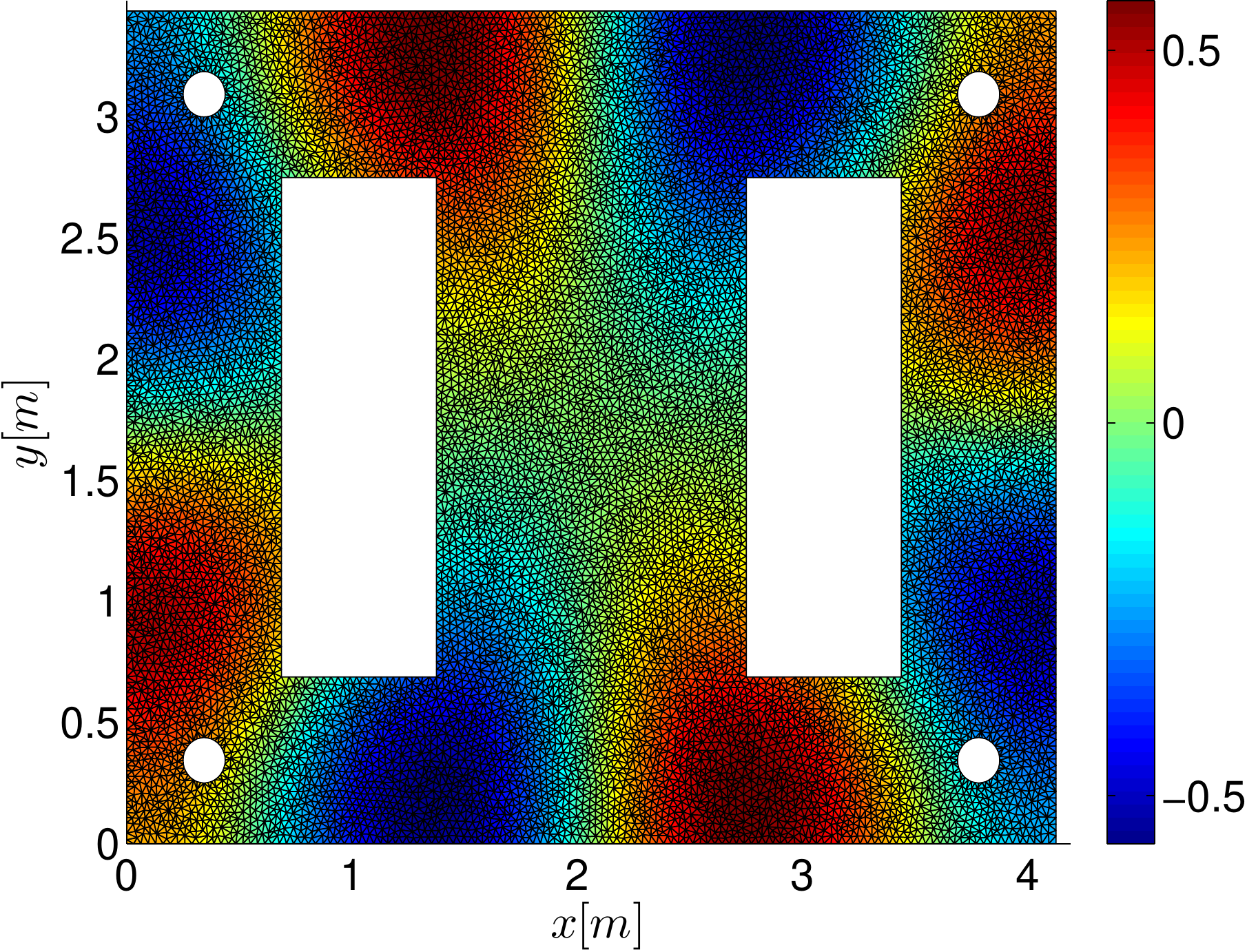}
       \caption{13th eigenfunction}
    \end{subfigure}
\begin{subfigure}[b]{7.3cm}
      \includegraphics[width=7.2cm]{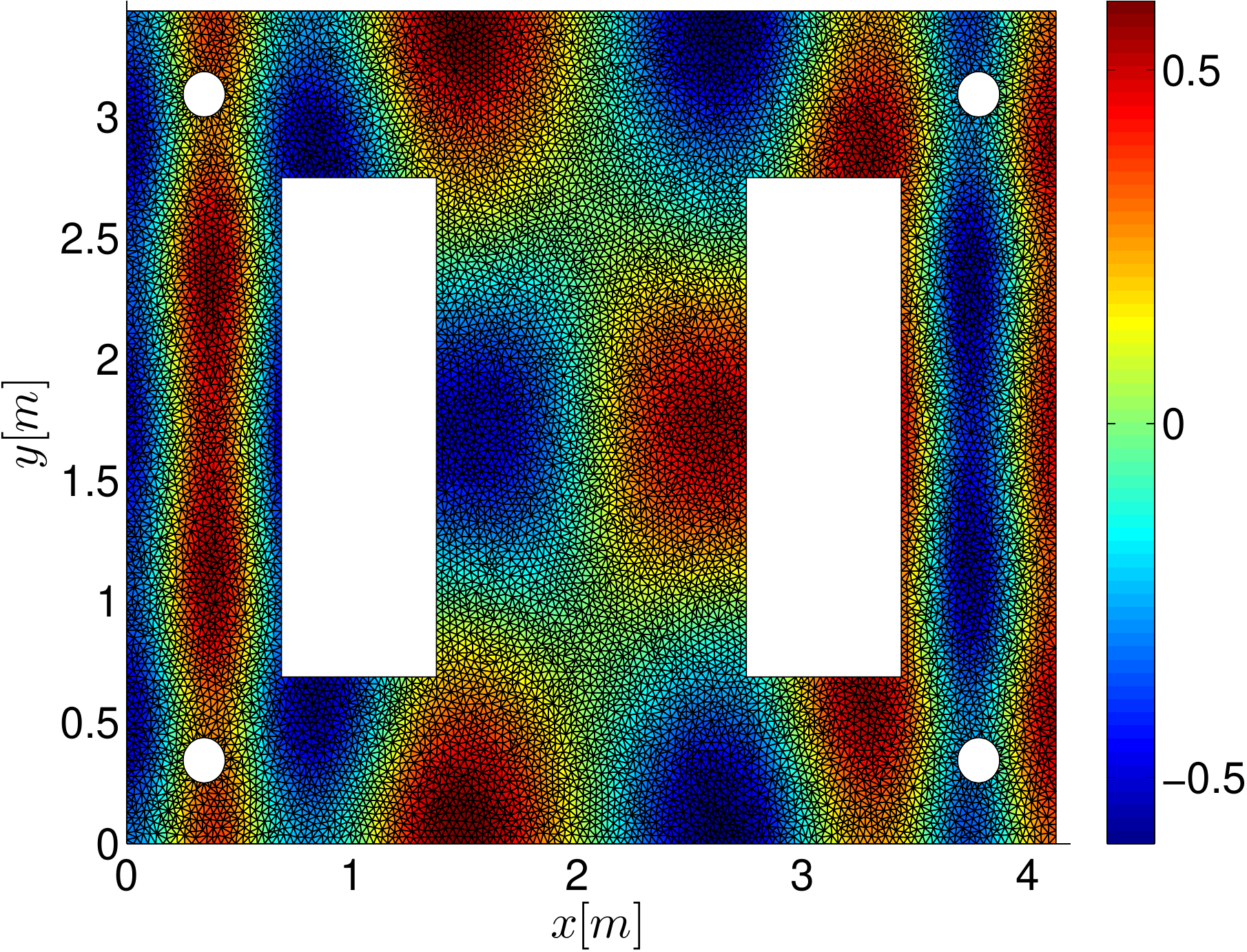}
       \caption{30th eigenfunction}
    \end{subfigure}
 \caption{Two of the lower eigenfunctions (a) 1st eigenfunction (b) 5th eigenfunction and two of the higher eigenfunctions (c) 13th eigenfunction and (d) 30th eigenfunction}
 \label{modes}
   \end{figure}

\begin{figure}[ht]
   \centering
    \begin{subfigure}{7.8cm}
     \label{d=2}
     \centering 
      \includegraphics[width=6.2cm]{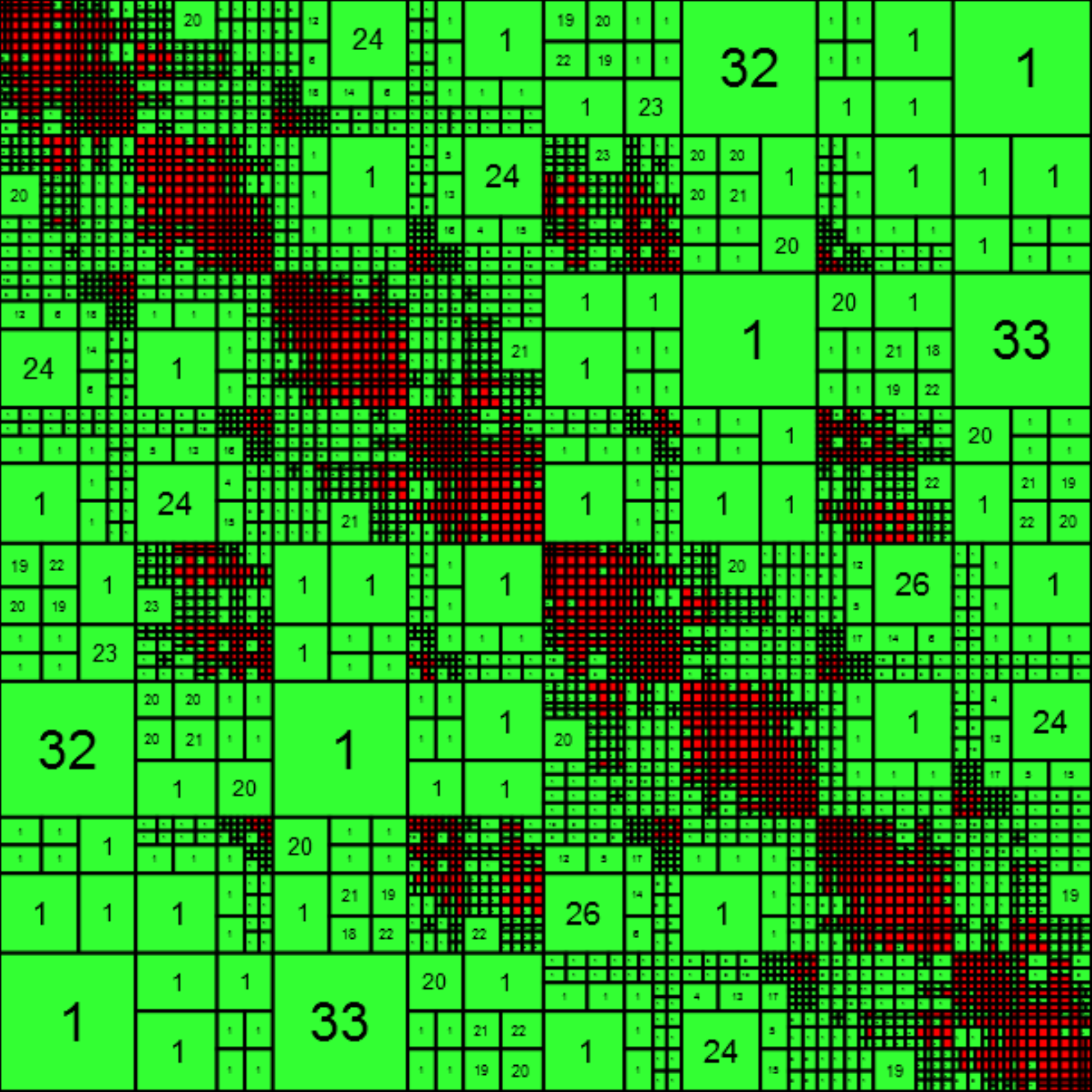}
       \caption{Hierarchical matrix $\tilde{\mathbf{{A}}}$ for $d=2m$}
    \end{subfigure}
\begin{subfigure}{6.8cm}
   \label{d100}
  \centering
      \includegraphics[width=6.2cm]{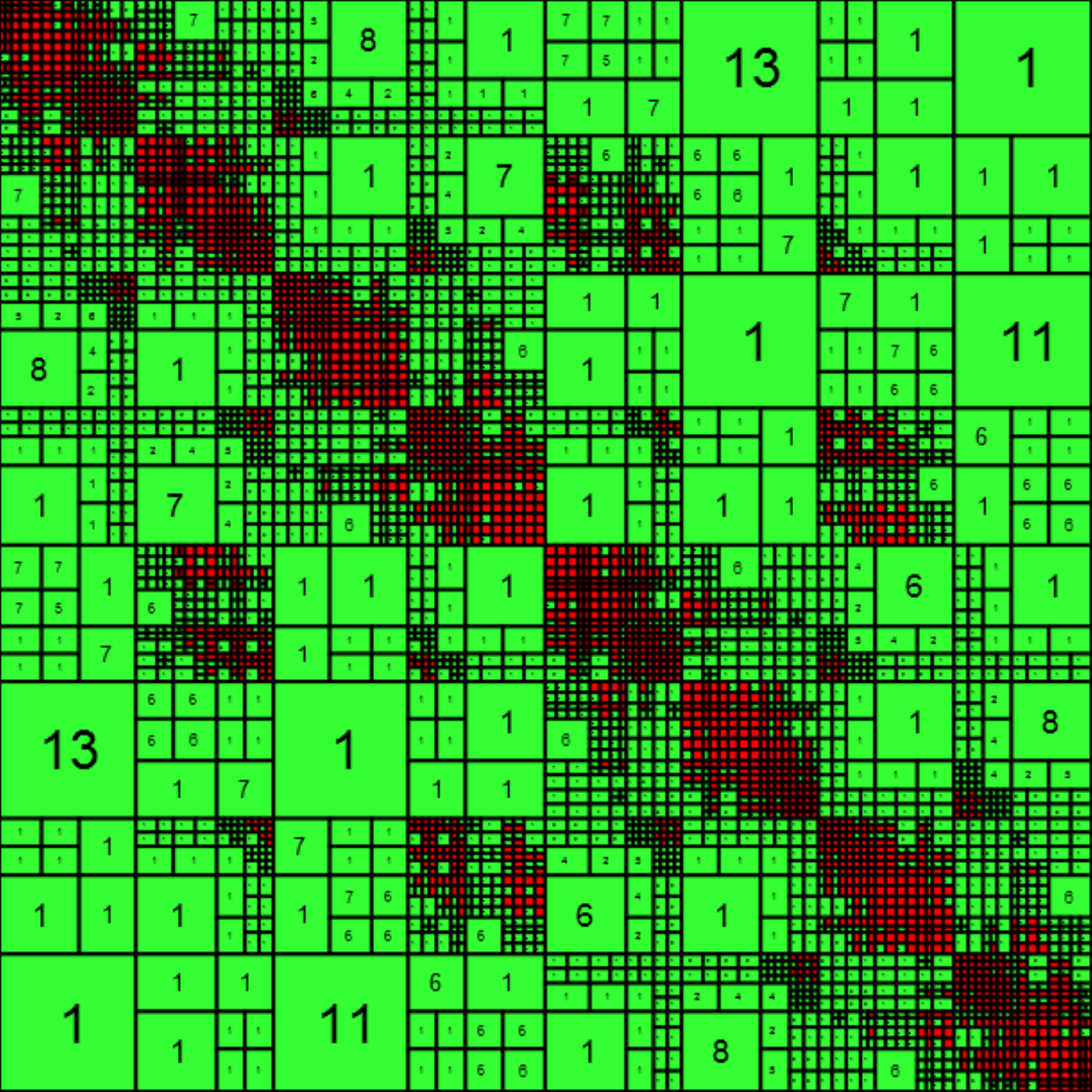}
        \caption{Hierarchical matrix $\tilde{\mathbf{{A}}}$ for $d=10m$}
    \end{subfigure}
 \caption{Visualisation of hierarchical matrices (a) for $d=2m$ and (b) for $d=10m$}
 \label{vizu}
   \end{figure}

Next the solution for the eigenfunctions is shown only for a correlation length of $d=2m$. Two of the lower eigenfunctions, such as the 1st and the 5th and two of the higher eigenfunctions, such as 13st and the 30th, are depicted in Fig.(\ref{modes}a) and Fig.(\ref{modes}b), respectively. The eigenfunctions carry information about the shape of the random realizations of the random field. Due to the decay of the eigenvalues the influence of the higher eigenfunctions is suppressed, hence, they are less significant. In the discrete version of the KLE, the eigenvectors (eigenfunctions in the continuous case) that contain most of the information are called principle components \citep{doi:10.1142/S021820250300257X}. 
In order to obtain a better insight into the structure of the hierarchical matrices used in the computations two of them are shown in Fig.(\ref{vizu}). The maximal rank of the admissible blocks in Fig.(\ref{vizu}a) and Fig.(\ref{vizu}b) is $k=33$ and $k=13$, respectively. This result suggests that for strongly correlated random fields the compression of the dense matrix becomes better.

The next section gives more detailed information about the actual memory storage benefits of using hierarchical matrices.

\begin{table}[ht]

  \begin{center}
  \begin{tabular}{| c | c || c | c || c | c |}
  \hline

  $N_{t}$ & $\mathbf{A}$ & $\tilde{\mathbf{{A}}}$($d=2m$) & $\Delta $ & $\tilde{\mathbf{{A}}}$($d=10m$) & $\Delta$ \\
  \hline
  
  $1320$  & \hspace{+0.085cm}$13$ MB & \hspace{+0.075cm} $11$  MB & $2,81 \cdot 10^{-4}$ & \hspace{+0.075cm}$11$ MB & $7,42 \cdot 10^{-4}$  \\
  \hline
  
  $7545$  & $434$ MB & $146$ MB & $2,37 \cdot 10^{-4}$& $139$ MB & $5,59 \cdot 10^{-4}$ \\
  \hline
  
  $24727$ & $4664$ MB (nem) & $593$ MB & $---$ & $556$ MB & $---$\\
  \hline
   
  $35450$  & $9588$ MB (nem) & $949$ MB & $---$& $887$ MB & $---$ \\
  \hline
    \end{tabular}
   
  \end{center}
     \caption{Memory requirements for the full format and hierarchical format of matrices.}
     \label{table1}
  \end{table}
 \subsection{Memory Storage Benefits}
Table \ref{table1} shows the memory requirements for storing the full matrix $\mathbf{{A}}$ and its hierarchical representation $\tilde{\mathbf{{A}}}$ and the relative error computed in the $l_{2}-$norm,
 \begin{equation}
	\label{errorr}
\Delta =\frac{\norm{\mathbf{A}-\mathbf{\tilde{A}}}_{l_{2}}}{\norm{\mathbf{A}}_{l_{2}}}
\end{equation}
for a different number of triangles in the mesh and different correlation lengths. For $N_{t}=1320$ the memory storage benefits are not significant, however, when $N_{t} \geqslant 24727$ the memory allocation fails on a $64$ bit desktop computer with 24 GB installed RAM memory (not enough memory - "nem") and the benefits are obvious. The memory issues are also reported in \citep{Khoromskij2009}. Table \ref{table1} shows that the compression of the dense matrix indeed becomes better when the correlation length is increased, as expected.
 
\subsection{Statistics of the Inductance}
\begin{figure}[t]
\captionsetup[subfigure]{justification=centering}
\centering
\hspace{0.3cm}
\begin{subfigure}{0.46\textwidth}  
\centering
\begin{tikzpicture}[scale=0.75,trim axis left, trim axis right]

     \begin{axis}[
        xlabel=\large {$d \ [m]$},
        ylabel=\large {$L_{\mu}\,\ [H]$}, x tick label style={rotate=0},yticklabel style={/pgf/number format/fixed,
                  /pgf/number format/precision=4},xlabel shift = 0 pt,ylabel shift =  -2pt ,
                         ]
      \addplot table{mean_value.txt};            
    \end{axis}
      \end{tikzpicture}
   
      \caption{\centering Mean value of the inductance}
      
\end{subfigure}
\begin{subfigure}{0.43\textwidth}  
\centering
\begin{tikzpicture}[scale=0.75,trim axis left, trim axis right]

     \begin{axis}[ 
        xlabel=\large {$d \ [m]$},
        ylabel=\large {$L_{std} \,\ [H]$},  x tick label style={rotate=0},yticklabel style={/pgf/number format/fixed,
                /pgf/number format/precision=5},xlabel shift = 0 pt,ylabel shift =  0 pt
                       ]
      \addplot table{standard_deviation.txt};
             
    \end{axis}

      \end{tikzpicture}
      
      \caption{\centering Standard deviation of the inductance}
     
      \end{subfigure}
     
      \caption{The mean value (a) and the standard deviation (b) of the inductance}
       \label{statistics}
     \end{figure}
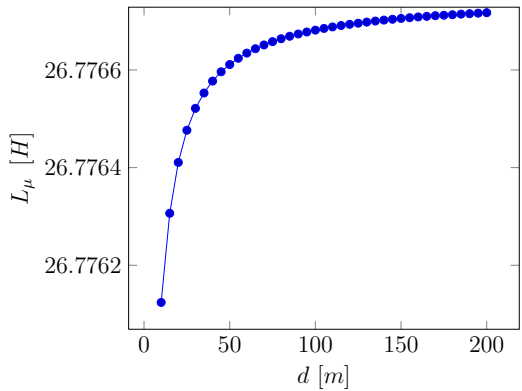
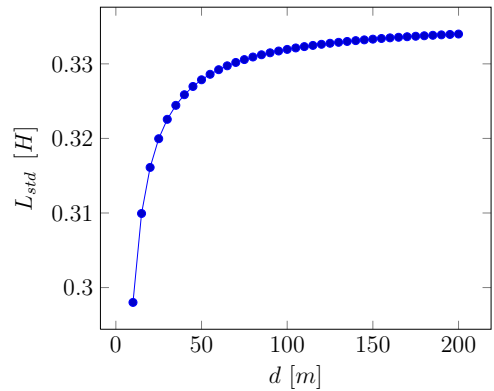

In this section we apply the KLE to study the mean value and the variance of the QoI. The covariance function of the random reluctivity is assumed to be the exponential function (\ref{exponential}) with a standard deviation $\sigma=10$. The mean value of the random field is assumed to be $\overline{\nu}=795.774$ $H^{-1}m$. Both, the air region and the coil region have the reluctivity of vacuum, $\nu_e=\nu_j=\mu_{0}^{-1}$. The number of turns in the primary coil is $N_{str}=260$. The random variables $\xi_{i}$ are assumed to be independent and uniformly distributed in the interval $[-\sqrt{3},\sqrt{3}]$. Equations (\ref{m_value}) and (\ref{vsr_value}) are solved by using the stochastic collocation method, with polynomial degree $p=2$ in each dimension, for different correlation lengths of the random reluctivity. The results are shown in Fig.(\ref{statistics}).
     
As it can be seen from Fig.(\ref{statistics}), both, the mean value and the standard deviation increase as the correlation length increases. However the changes in the mean value due to the increasing correlation length seem to be negligible compared to those of the standard deviation. After a certain value of the correlation length the statistics of the inductance is no longer influenced by its increase. This result is expected, as the covariance function becomes almost constant across the core region as it can be seen from the following limit:
 \begin{equation}
	    \label{limit}
		  \lim_{d\to\infty}\sigma^{2}\exp{\Bigg( -\frac{||\vec{x}-\vec{y}||_{l_{1}}}{d}\Bigg)}=\sigma^{2}.
	\end{equation} 
The covariance between every two points in space approaches the (constant) variance of the random field

\section{Conclusion and Outlook}

An efficient approach for computing the KLE based on the Lanczos algorithm and the hierarchical matrix technique has been presented in this paper. The KLE has been computed in a domain where the core of a single phase transformer is located. The computational benefits of using the hierarchical matrix technique are illustrated via memory storage benefits. The KLE is further used to determine the statistics of the inductance of a single phase transformer as a QoI. The mean value and the variance are calculated for different correlation lengths of the random reluctivity. The suggested approach is suitable for the stochastic modelling of strongly correlated random fields due to the fast decay of the eigenvalues. Future work on this topic will be related to the stochastic modeling anisotropic magnetic materials.

\end{document}